\documentclass[]{spie}  %>>> use for US letter paper
\usepackage{authblk}

\usepackage{amsmath,amsfonts,amssymb}
\usepackage{graphicx}
\usepackage{aas_macros}
\usepackage[colorlinks=true, allcolors=blue]{hyperref}

\title{Observing exoplanets from Antarctica in two colours: Set-up and operation of ASTEP+ }

\author[a]{Fran\c{c}ois-Xavier Schmider} 
\author[a]{Lyu Abe}
\author[a]{Abdelkrim Agabi}
\author[a]{Philippe Bendjoya}
\author[c,d]{Nicolas Crouzet}
\author[b]{Georgina Dransfield}
\author[a]{Tristan Guillot}
\author[a]{Olivier Lai}
\author[a]{Djamel Mekarnia}
\author[a]{Olga Suarez}
\author[b]{Amaury H.M.J. Triaud}
\author[a]{Philippe Stee}
\author[d]{Maximilian N. G\"unther}
\author[d]{Dennis Breeveld}
\author[d]{Sander Blommaert}
\affil[a]{Universit\'e C\^ote d'Azur, Observatoire de la C\^ote d'Azur, CNRS, Lagrange Laboratory, Nice, France}
\affil[b]{School of Physics \& Astronomy, University of Birmingham, Edgbaston, B15 2TT, Birmingham, UK}
\affil[c]{Leiden Observatory, Leiden University, Postbus 9513, 2300 RA Leiden, The Netherlands}
\affil[d]{European Space Agency (ESA), European Space Research and Technology Centre (ESTEC), Keplerlaan 1, 2201 AZ Noordwijk, The Netherlands}

\authorinfo{Further author information: Send correspondence to F.X. Schmider (schmider@oca.eu)}
\begin{document} 
\maketitle
%\title{Set-up and operation of a new two-colour camera at the ASTEP telescope in Antarctica}

\begin{abstract}
%A new two-color photometer has been successfully installed on the ASTEP telescope at the Concordia station in Antarctica, with the goal of high-precision photometric follow-up of exoplanetary transits. With this new camera, the ASTEP+ telescope will ensure the follow-up and the characterization of a large number of exoplanetary transits in the coming years in view of the future space missions JWST and Ariel.
On December 2021, a new camera box for two-colour simultaneous visible photometry was successfully installed on the ASTEP telescope at the Concordia station in Antarctica. The new focal box offers increased capabilities for the ASTEP+ project. The opto-mechanical design of the camera was described in a previous paper\cite{2020SPIE}. Here, we focus on the laboratory tests of each of the two cameras, the low-temperature behaviour of the focal box in a thermal chamber, the on-site installation and alignment of the new focal box on the telescope, the measurement of the turbulence in the tube and the operation of the telescope equipped with the new focal box. We also describe the data acquisition and the telescope guiding procedure and provide a first assessment of the performances reached during the first part of the 2022 observation campaign. Observations of the WASP19 field, already observed previously with ASTEP, demonstrates an improvement of the SNR by a factor 1.7, coherent with an increased number of photon by a factor of 3. The throughput of the two cameras is assessed both by calculation of the characteristics of the optics and quantum efficiency of the cameras, and by direct observations on the sky. We find that the ASTEP+ two-colour transmission curves (with a dichroic separating the fluxes at 690nm) are similar to those of GAIA in the blue and red channels, but with a lower transmission in the ASTEP+ red channel leading to a 1.5 magnitude higher B-R value compared to the GAIA B-R value. With this new setting, the ASTEP+ telescope will ensure the follow-up and the characterization of a large number of exoplanetary transits in the coming years in view of the future space missions JWST and Ariel.

\end{abstract}


\section{INTRODUCTION}
\label{sec:intro}  % \label{} allows reference to this section\
% ASTEP has been developed at Dome C for search of exoplanetary transits\cite{Fressin+2007, Daban+2010, Guillot+2015}.
ASTEP (Antarctic Search for Transiting ExoPlanets) has been developed to search for exoplanetary transits from Dome C\cite{Fressin+2007, Daban+2010, Guillot+2015}.
It has been able to produce high-precision photometric observations, both for detection of new candidates and for the follow-up of already known exoplanets\cite{Bouma+2020AJ, Dawson+2021, Dong+2021, Grieves+2021, Burt+2021, Kaye+2022, Wilson+2022, Mann+2022, Christian+2022, Dransfield+2022}. The location of ASTEP in Antarctica makes it unique for the follow-up of long duration transits and very southern objects detected by TESS and in the continuous viewing zones of JWST and Ariel.
In 2020, a new focal box was developed at the Lagrange laboratory with the support of ESA/ESTEC and the University of Birmingham. The support permitted the acquisition of two cameras for two-color (blue and red parts of the visible light) simultaneous photometry. The focal box design and its characteristics are described in a previous SPIE paper \cite{2020SPIE}.

After the development and tests at the laboratory, the new focal box was shipped and installed at the Concordia station, Dome\,C in Antarctica. The new ASTEP+ telescope has started its observations in March 2022 for the whole winter season.

Here we describe the different steps of the assembly and tests of the focal box, its shipment to Dome\,C, its installation on ASTEP at Concordia during the austral summer season, as well as its operation during the first months of the polar winter season and the results achieved. In a parallel paper \cite{2022SPIE+}, we describe the target selection process  and the data automatic reduction, in view of a fully robotic observation facility.

\section{FOCAL BOX ASSEMBLY AND LABORATORY TESTS}
 
 Before shipping the focal box to the Concordia station, it as necessary to assemble it at the laboratory and test its behaviour for the expected environmental conditions in Antarctica, where the external temperature can vary from -20$^\circ$C down to -80$^\circ$C. These conditions impose a special conception of the box in different compartments with different functions, an external isolation and internal temperature regulation to ensure the correct working properties of the focal box. 
 
\subsection{Assembly}
\label{sec:assembly}

In the previous paper \cite{2020SPIE}, we described the optical conception and the realisation of the different mechanical supports for the optics and the cameras. The main carbon composite plate and all opto-mechanical modules of the new focal box were assembled on an optical bench in the laboratory to align the optics. In addition to the mechanical assembly, several heating elements were added in the different compartments of the focal box, so we could guarantee different working temperature following the requirements of each component. The optical compartment could remain colder than 0$^\circ$C, when the cameras and their electronics have to stay in a positive temperature environment. Thermal insulation was added all around the box and between each compartment. The temperature control of the different components is carried out by several stand-alone regulators with PID algorithm using PT100 probes.

 \begin{figure}
\begin{center}
\begin{tabular}{c}

\includegraphics[height=5.5cm]{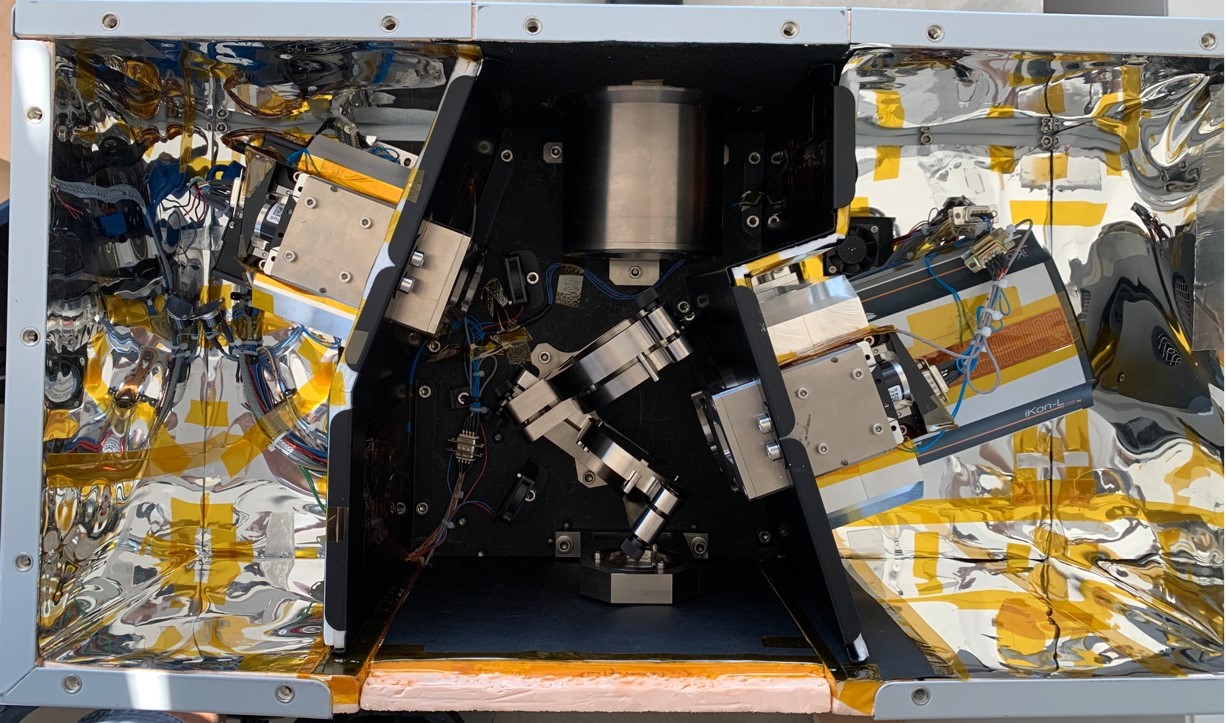}
\includegraphics[height=5.5cm]{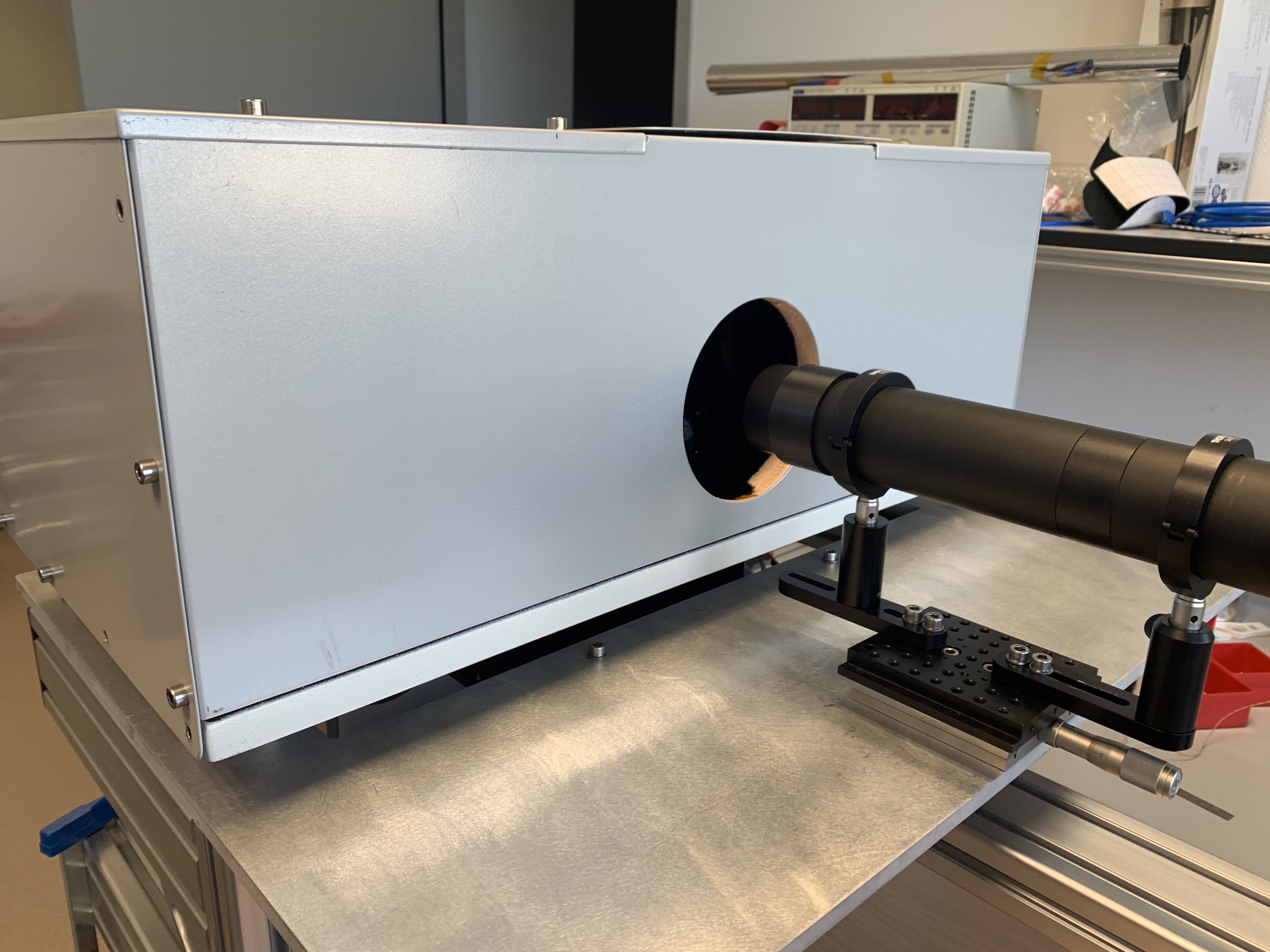}
\end{tabular}
\end{center} 
\caption  {The new focal box assembled in the lab for alignment with an artificial optical source (right) \label{fig:labtests} } 
\end{figure}

\subsection{Laboratory tests}
\begin{figure}
\begin{center}
\begin{tabular}{c}

\includegraphics[height=5.5cm]{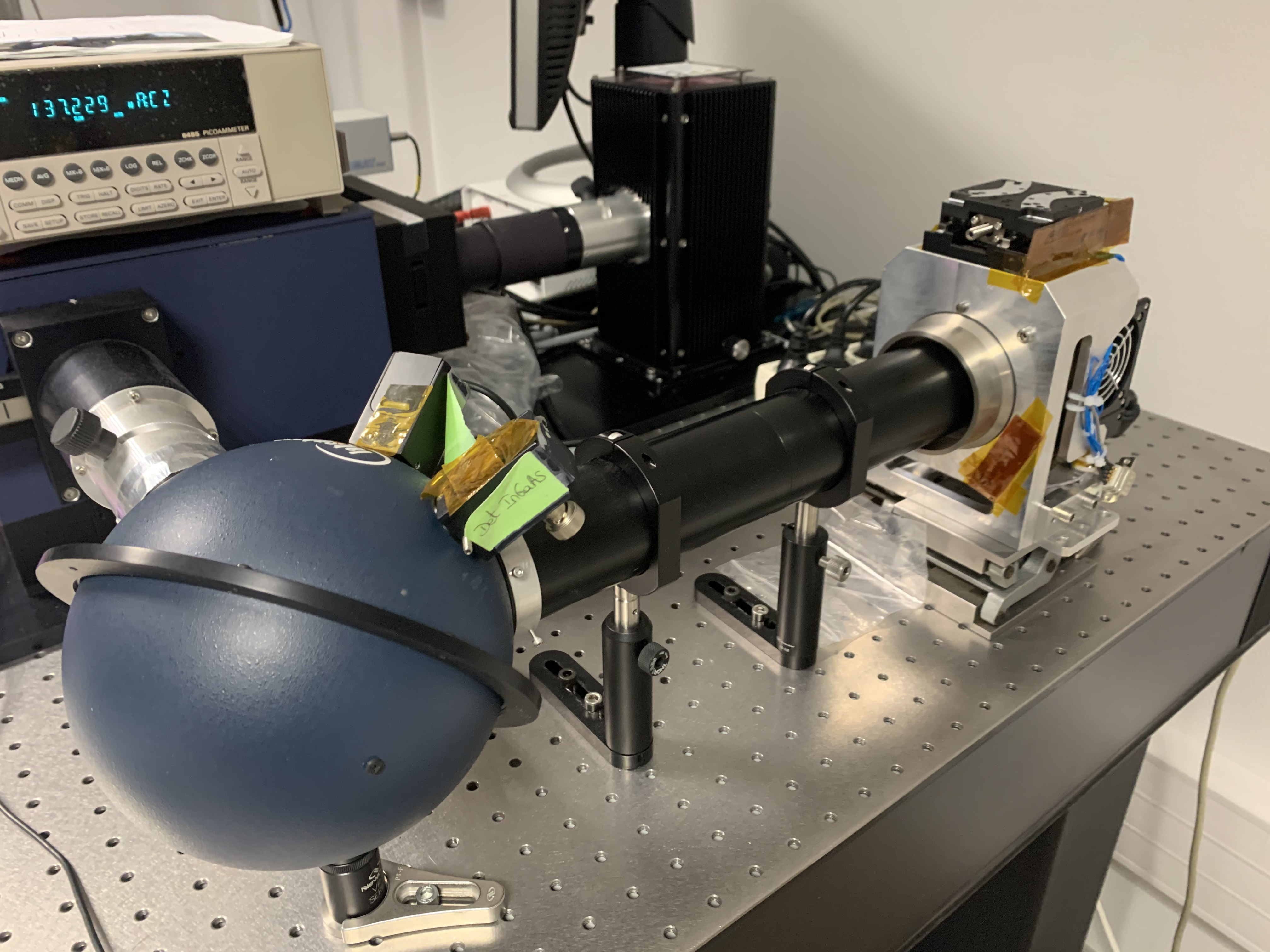}
\end{tabular}
\end{center} 
\caption  {FLI camera test behind the optical source \label{fig:labtests1} } 
\end{figure}

\begin{figure}
\begin{center}
\begin{tabular}{c}

\includegraphics[height=5.5cm]{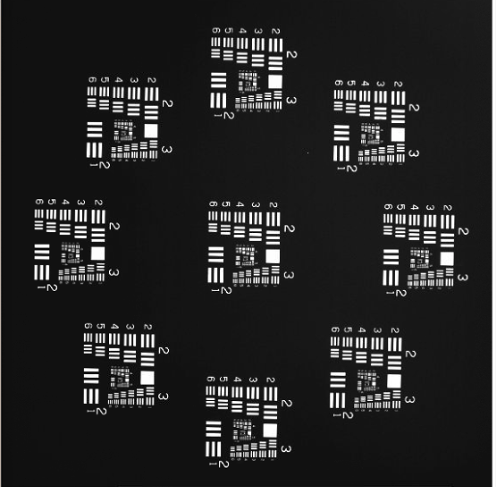}
\includegraphics[height=5.5cm]{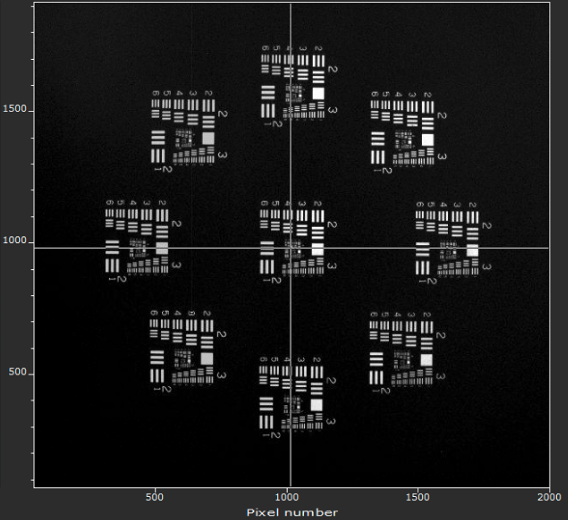}
\end{tabular}
\end{center} 
\caption  {Optical pattern as seen by the FLI \& ANDOR cameras\label{fig:labtests2} } 
\end{figure}

% \textcolor{red}{Karim, Lyu}
At the laboratory, several tests were carried out on each element before assembly and set-up. Each camera was placed on an optical bench with a monochromator and an integrating sphere (Figure \ref{fig:labtests1}) in order to estimate the Pixel Response Non Uniformity (PRNU), the linearity of the detector and the read-out noise level. The whole box was then aligned on an external light source and the quality of the image was verified using an optical pattern (Figure \ref{fig:labtests2}).

A first adjustment of the opto-mechanical modules consists of aligning all the elements with a laser and, at the entrance of the box, of positioning a grid located in a tube (telescope simulator) at the theoretical focal plane of the telescope. The alignment of the optical axes of the two cameras with the optical axis of the telescope is obtained by centering the image of the pattern on each camera, which is done thanks to the adjustment of the dichroic beamsplitter for the blue channel and of the mirror for the red channel.


The optical transmission  of the new focal box (Fig. \ref{fig:transmissionASTEP}) was estimated taking into account the cameras QE, the transmission and reflection of the dichroïc beamsplitter and the coating of the lenses and of the mirrors. The primary (M1) and secondary (M2) mirrors of the telescope are coated with protected aluminium, which has a gap in reflection around 800 nm, explaining the gap in the transmission of the red channel. The central wavelength of the blue channel is 555 nm, with a FWHM of 293 nm, while the central wavelength of the red channel is 850 nm, with a FWHM of 276 nm. In the future, we foresee to change the coating of M2 mirror, inclined at 45 degrees, by a protected silver coating to increase the total transmission efficiency of the instrument. %We found that the ASTEP transmission of the blue and red channels are almost identical to the GAIA blue and red channels respectively. The red channel is although slightly redder with a reduced transmission than the GAIA one. 
In the section \ref{sec:results}, we show the response of the focal box as compared to GAIA for all measured stars in a given field.

The thermal behaviour of the box was tested in a thermal chamber. The external temperature was varied to simulate the conditions at Dome C both in summer and in winter, with a temperature range between -20$^\circ$C and -80$^\circ$C.
The lower part of the camera box contains electronic modules, in particular the Andor camera power supply control electronics and a converter that allows the USB signal to be transferred from the cameras to the acquisition computer. These electronic devices produce heat when cameras are in function.
This part of the box is also thermalized using dedicated temperature controllers. Fans installed in the compartment of each camera produces heat exchange between these compartments and, thus, maintain the whole at a set temperature around 10$^\circ$C.
Tests in the thermal chamber were carried out which showed that 50\% of the heating power was necessary to guarantee the good functioning of the whole installation in extreme cases and 80\% during the periods when scientific cameras are not dissipating heat (off mode).

 \begin{figure}
\begin{center}
\begin{tabular}{c}
\includegraphics[height=5cm]{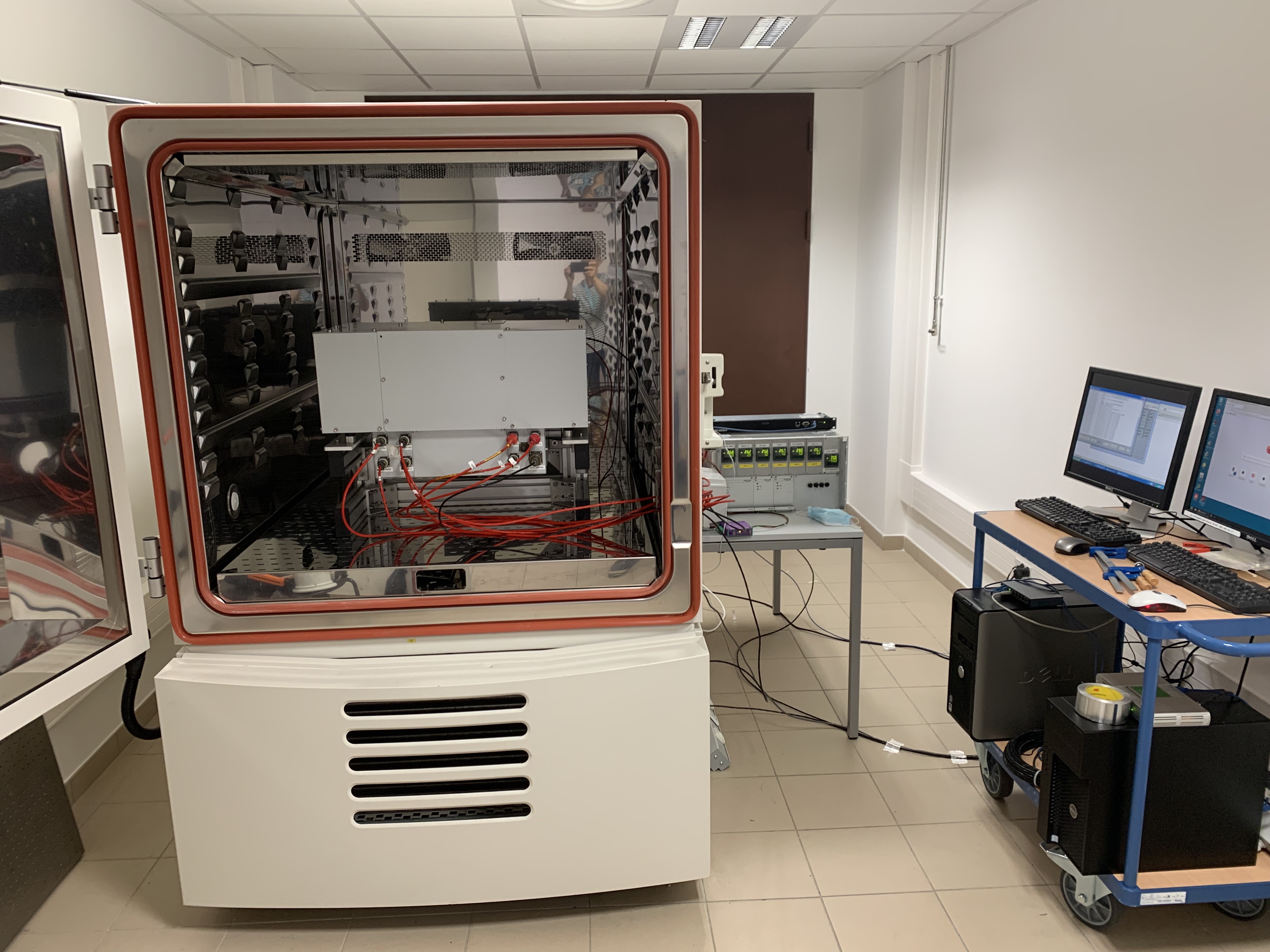}
\includegraphics[height=5.5cm]{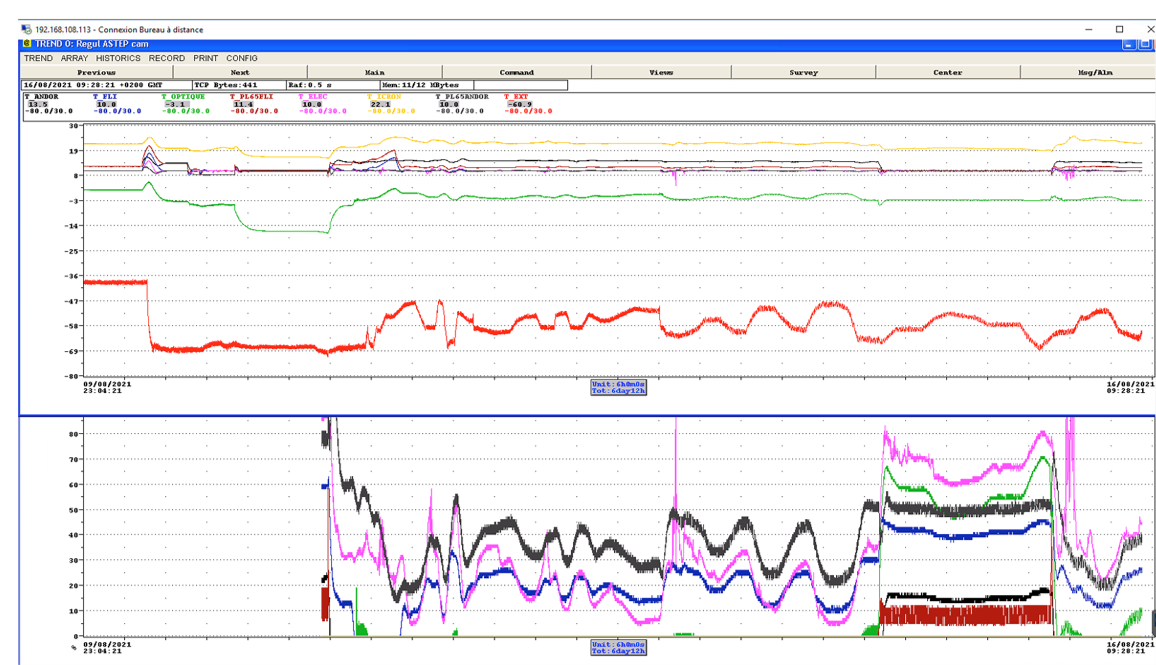}
\end{tabular}
\end{center} 
\caption  {Left: The new focal box is placed inside a thermal chamber, able to reach a temperature of -75$^\circ$C, similar to that of the Antarctica winter. The graphics on the right displays the temperatures measured at different places inside the box obtained when varying the environment temperature. \label{fig:ThermalTests} } 
\end{figure}

\section{SET-UP and OPERATIONS}
%\textcolor{red}{Karim, Djamel}
 
\subsection{Installation}
The new box was shipped to Dome\,C on October 5th, 2021 and was installed on the telescope during the austral summer campaign. The summer campaign was organised between November 23rd and January 25th by the ASTEP team composed of Karim Agabi, Djamel Mekarnia, Georgina Dransfield and Olivier Lai.  %\textcolor{red}{TBC}

%%\begin{table}[ht]
%%\caption{Main dates} 
%%\label{tab:calendar}
%%\begin{center}       
%%\begin{tabular}{|l|l|} %% this creates two columns
%% |l|l| to left justify each column entry
%% |c|c| to center each column entry
%% use of \rule[]{}{} below opens up each row
%%\hline
%%\rule[-1ex]{0pt}{3.5ex}  Date & Action \\
%%\hline
%%\hline
%%\end{tabular}
%%\end{center}
%%\end{table} 
%%

 \begin{figure}
\begin{center}
\begin{tabular}{c}
\includegraphics[height=9cm]{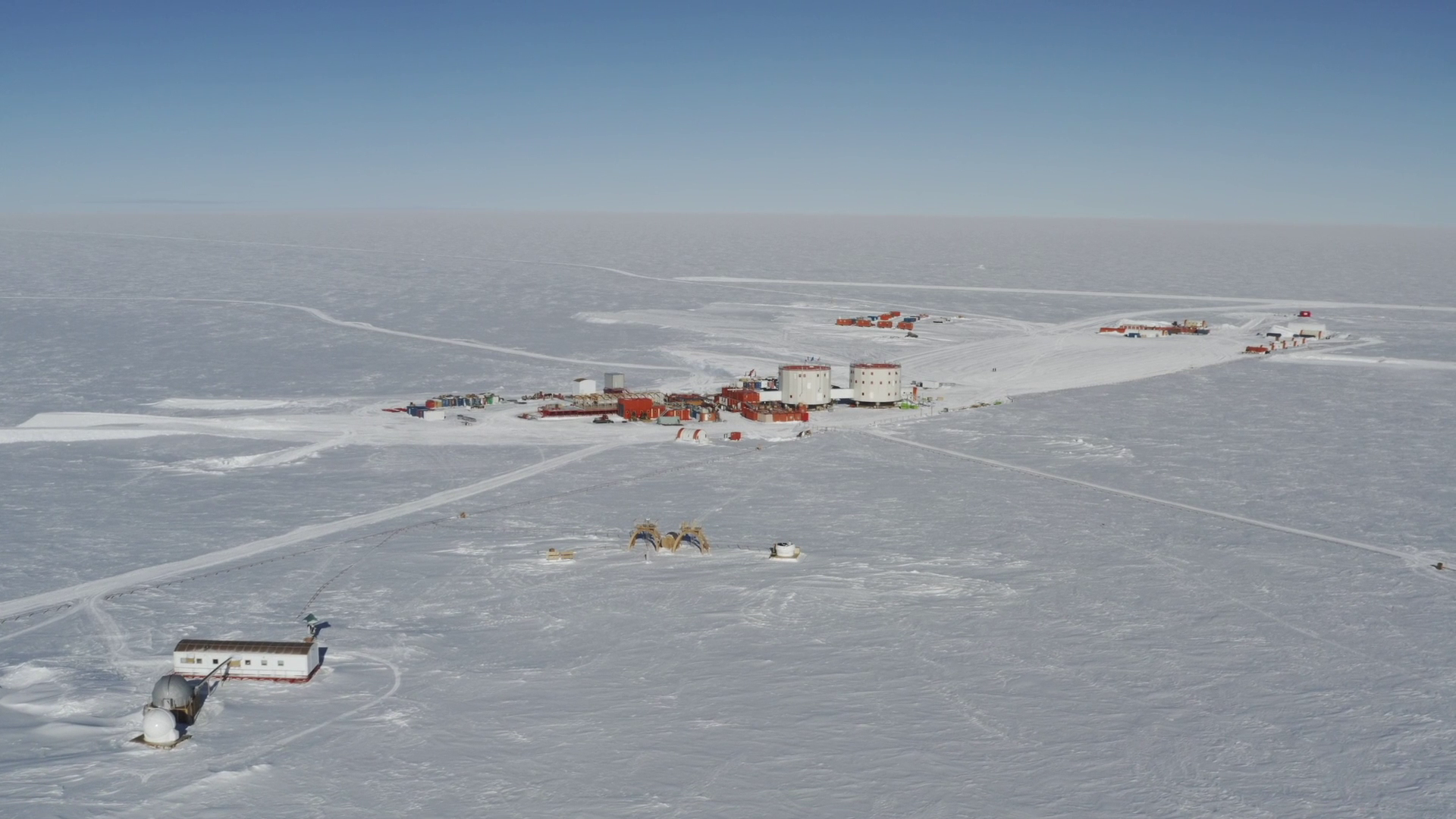}
\end{tabular}
\end{center} 
\caption  {Aereal view of the Concordia station. The two main buildings of the base are located in the center, while the summer camp is in the background of the picture. The new location of ASTEP telescope is the white Baader dome located in the foreground close to the Astro shelter. \label{fig:astroarea} } 
\end{figure} 
%textcolor{red}{New dome, transport of the telescope, installation and tests of the focal box
%Include picture there.}
%The new box was transported to Dome C in autumn 2021 for an installation during the summer campaign 2021-2022. 
 
 During the summer campaign, we seized the opportunity to move the telescope to a new location on the base, in the so-called Baader dome. This displacement, already scheduled and prepared one year before, aims at grouping the French and Italian astronomical activities in the same laboratory (called Astro shelter). The new dome is better suited for the ASTEP+ telescope as it offers more space and an automatic closure capability. In addition to this huge operation, we also removed all ASTEP experiment elements from their initial installation to their new location: electronics devices were installed close to the new Dome and the telescope control computers were installed inside the Astro shelter. We also installed a new primary mirror on the telescope, a new data-processing server, conducted a whole maintenance of the telescope and its mount and updated all scripts linked to the data acquisition and data processing.

\begin{figure}
\begin{center}
\begin{tabular}{c}
\includegraphics[width=9.cm]{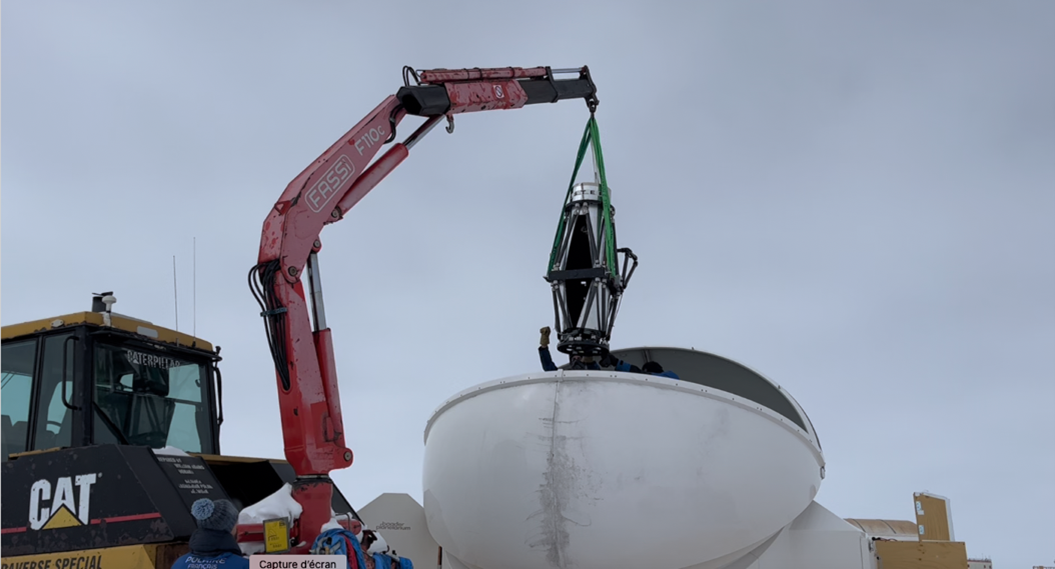}
\hskip 0.5cm
\includegraphics[width=6.5cm]{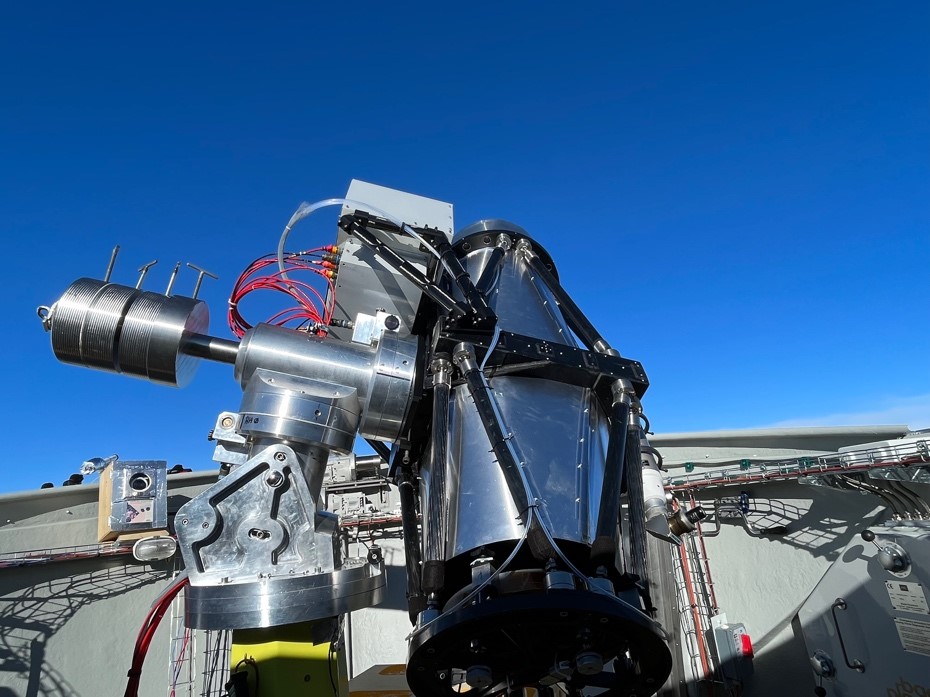}
\end{tabular}
\end{center} 
\caption  {Left: Installation of the ASTEP+ telescope in its new dome, using the logistical means of the station. Right: The ASTEP+ telescope equipped with its new focal box fully installed in the new dome. \label{fig:Baader} } 
\end{figure} 

After the displacement of the telescope, the new box was placed on the telescope. The optical alignment of the focal box was first checked at the laboratory, using an internal light source, before its installation on the telescope. Then, the alignment of the telescope with the focal box was done by pointing a bright star, usually Canopus, visible in daylight, thanks to the exceptional transparency of the sky at Dome\,C. A density filter was added on each camera to avoid saturation by the daylight. The optical adjustment procedure consists in aligning the focal box optical axis with the telescope axis, which supposed to move the M1 mirror on three axis X,Y, and Z. Then, the alignment of the equatorial mount was checked to ensure a perfect guiding on the field for several hours. We cannot reach an alignment better than a few arc minutes due to the small number of stars that can be observed during this period of continuous daylight, and of the absence of a pointing model taking into account the flexion of the telescope. This set-up is generally good enough to ensure the operations, as the ASTEP+ field is about 40 arc minutes wide, and a field recognition is done for each new observation. Real-time guiding is also ensured from the image of the blue camera. However, for fields close to the Celestial Pole, the uncertainty of the mount set-up results in a noticeable field rotation, reducing the photometric quality of stars at the edges of the image. 
%, a little constraining for our purpose. \textcolor{red}{Lyu: precision of set-up, field rotation at the pole ?} 
In the future, we expect to have a new more reliable mount, and to use an automatic set-up procedure to precisely align the mount to within a fraction of arc minutes.

\subsection{Turbulence measurement}

To prevent frost from forming, the telescope primary and secondary mirrors are heated from their back surface, as well parts of the camera focal box. This is necessary although it is well known that heat sources in optical beam are harmful to image quality. We ran a series of test in daytime to try to quantify the effect of this heating on the turbulence inside the tube and the dome, using a optical turbulence sensor, AIRFLOW \cite{Lai2019}. This instrument measures optical path differences along parallel beams within a sensing cell by interferometric means, and by plotting the phase variance as a function of beam separation, the phase structure function can be estimated. A value of $C_n^2$ can be extracted if the phase structure function follows the Kolmogorov structure, $D_{\phi}=(D/r_0)^{5/3}$.

\begin{figure}
\begin{center}
\begin{tabular}{c}
\includegraphics[height=7.5cm]{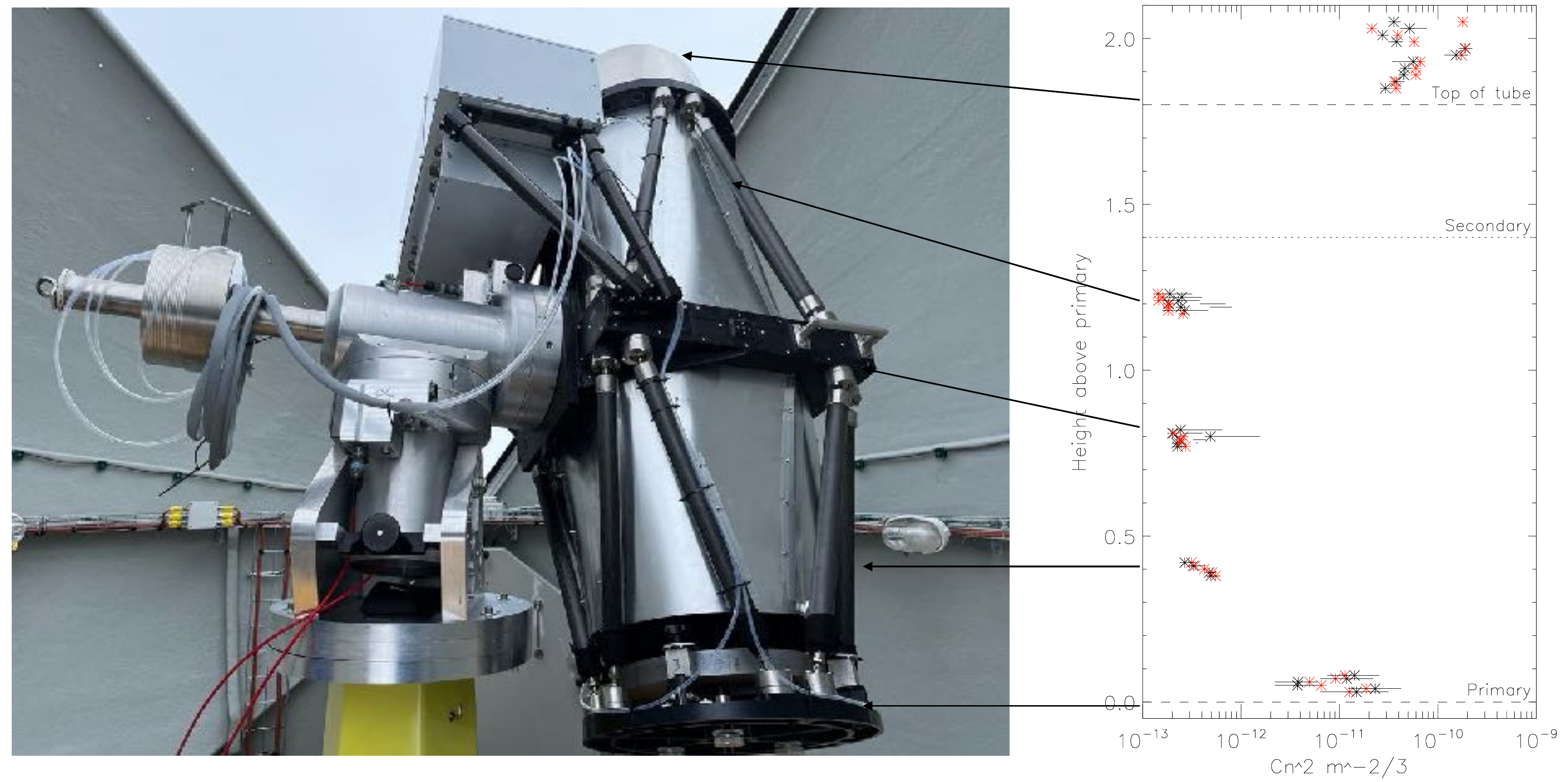}
\end{tabular}
\end{center} 
\caption  {Measurement of $C_n^2$ along the telescope tube in daytime. The primary mirror is not heated but the mixing of air at different temperatures around the air gap at the base of the tube increases the local turbulence. \label{fig:tubeseeing} } 
\end{figure}

There is a small air gap at the bottom of the tube, allowing air to flow around the primary mirror; without any active heating (but thermal driving from sunlight) we found that the air inside the tube was very calm but some mixing close due to this gap increased the turbulence around the primary mirror (Figure~\ref{fig:tubeseeing}). This illustrates that daytime tests are of limited operational value, although they can be useful to provide guidelines in this extreme environment.

\begin{figure}
\begin{center}
\begin{tabular}{c}
\includegraphics[height=6.5cm]{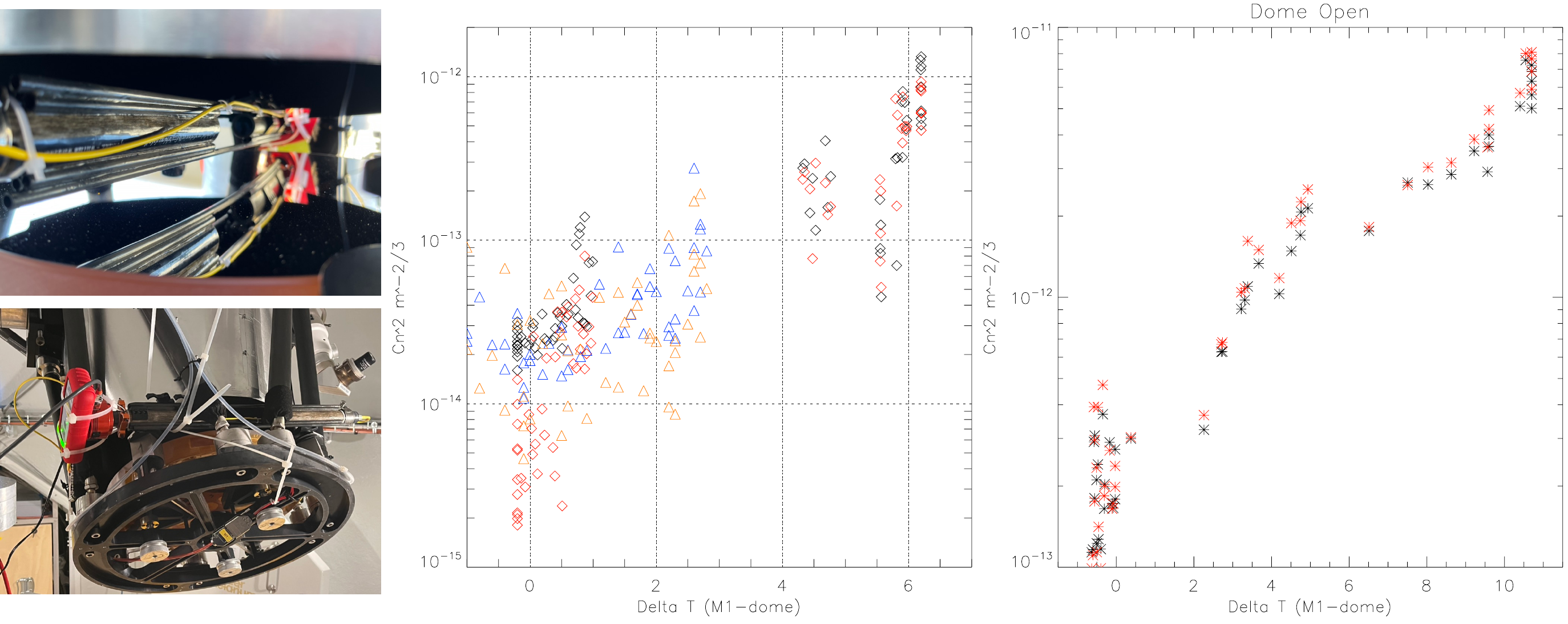}
\end{tabular}
\end{center} 
\caption  {$C_n^2$ as a function of the temperature difference between the primary mirror and the ambient air, measured just above the primary mirror (left). When the dome is closed (middle), there is exquisite sensitivity to $\Delta$T, while when the dome is closed (right), $C_n^2$ remains low for small $\Delta$T, presumably due to flushing, and a transition to turbulence occurs near $3^{\circ}C$. \label{fig:cn2deltaT} } 
\end{figure}

We ran turbulence measurements during mirror heating with the dome both closed and open (Figure~\ref{fig:cn2deltaT}). When the dome is closed, the environment around the telescope is very stable and we are very sensitive to minute heating. However, when the dome is open, which is much more representative of operational conditions, even low wind clears out some heat and we need to increase the temperature difference between the primary mirror and the ambient air to be able to measure the turbulence. Unfortunately, when the dome is open, sunlight differentially heats parts of the dome (we tried to prevent direct sunlight hitting the telescope during those tests), contributing to local turbulence.

Nonetheless we find that there is a regime (for $\Delta T < 3^{\circ}C$) where the air gap allows to flush the turbulence above the primary. The validity of these experiments is limited in day time, and a ruggedized version of AIRFLOW could be deployed in future observing seasons to provide quantitative control of the primary mirror temperature.

\subsection{Operations}

The observation software was inherited from the previous version of the instrument. Thanks to the modularity of the adopted software solution, the two new 2K$\times$2K backside illuminated cameras (ANDOR iKon-L 936 and FLI Kepler KL400) could be implemented rather easily. The main difference with the previous version is that the blue channel KL400 camera is used both as a guiding and a science detector. Short exposures (ranging from 0.1 to 5 seconds) are stacked to form $\sim$\,1 minute exposures. The particular sensor operation of the KL400 allows digitizing non-destructively the same image with two independent amplifier gain values using a 12 bits quantization (Low and High gains). The idea behind this operation mode being to have a finer sampling of lower intensities and a coarser sampling for higher intensities. FLI uses a somewhat complex procedure to calibrate these two images and stitch them together to form a full, 16 bits-equivalent image. In practice this stitching procedure was found too complex to be implemented on ASTEP+, and the non-linearity of each image resulted in non-satisfactory results for photometric purposes. Instead, a custom solution relying on the "linearization" of each frame using calibrated look-up-tables was envisaged but not yet implemented. In the current version of the software, only the high gain stacked images are used as science frames. The low gain images are used for guiding on brighter nearby stars when possible since they are less prompt to too high background levels situations, and the high gain images are used when the single-frame exposure time needs to be very short (below 2 seconds).

\subsection{Guiding}
% \textcolor{red}{LYU Communication issues}
In the previous version of our instrument, a guiding camera was used solely for guiding while ASTEP+ exploits the fast readout capability of the FLI KL400 camera to guide and save stacked science frames. Due to the mount inertia and the mechanical backlash, the guiding frequency can hardly be higher than 1 Hertz. Based on tests performed at the beginning of the polar winter season we set the guiding corrections frequency to 0.5\,Hz. Typically when the target star is brighter than V$\sim$11, the frame cadence is increased, but we keep the mount guiding corrections to be below or equal to 0.5\,Hz. Unfortunately, with increasing nighttime duration and temperatures going down below -50$^{\circ}$C we experienced random and too frequent communication losses between the acquisition PC and the instrument. After thorough investigations, we identified the USB3-to-Ethernet converter cable as the most probable cause of these failures. The distance between the telescope and the laboratory where the acquisition computer is hosted was about 60m, within the capability of the USB-Ethernet converter. Disgracefully, the frost imposes special cables able to resist to extreme temperature and the one we had was not sufficient for the required specifications of the USB3 norm on this distance. To avoid frequent interruptions in the acquisition, we decided to operate only the red camera, and to update the guiding from the science images at a least frequent rate. This proved to be efficient, and we did not noticed degradation of the photometric accuracy.

An upgrade of this cable is foreseen for the next instrument update campaign. In the meantime, we are investigating software solutions to mitigate the data losses caused by frequent communication failures. Note that this sometimes also affects the red Andor camera, but at a much lower rate (this camera uses an USB2 protocol). In the future, the new direct-drive mount will allow more frequent guiding updates which would also increase the precision.

\subsection{Focusing}
% \textcolor{red}{TBC LYU}
The focus of the images on the two detectors can be adjusted independently thanks to a movable lens placed just before the camera. During the first tests, the focus proved to be very sensitive to the external temperature, due to the thermal dilatation of the telescope parts. The focus is determined manually from the quality of the images and is adjusted regularly when the temperature changes. An empirical temperature dependency law has been determined. In the future, we foresee to set-up an automatic focus adjustment procedure.

\section{FIRST RESULTS}

\label{sec:results}

\begin{figure}
\begin{center}
\begin{tabular}{c}
\includegraphics[width=12.5cm]{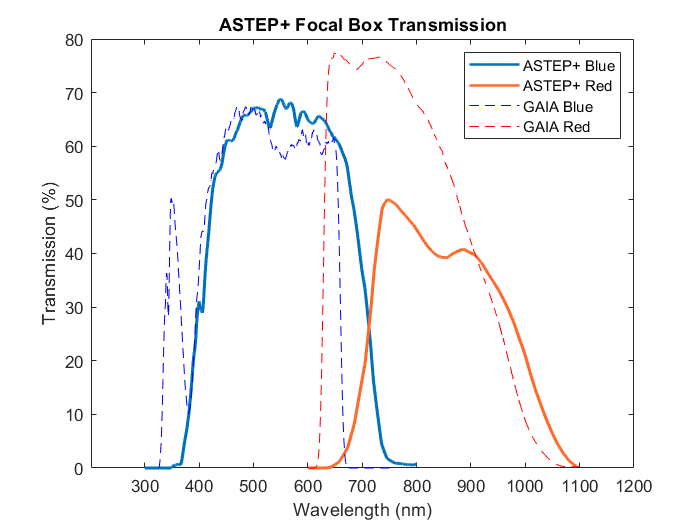}
\end{tabular}
\end{center} 
\caption  {Transmission of the two channels of the new focal box calculated from theoretical transmission curves as compared to the GAIA blue and red channels. \label{fig:transmissionASTEP} } 
\end{figure}

\begin{figure}
\begin{center}
\begin{tabular}{c}
\includegraphics[height=6.2cm]{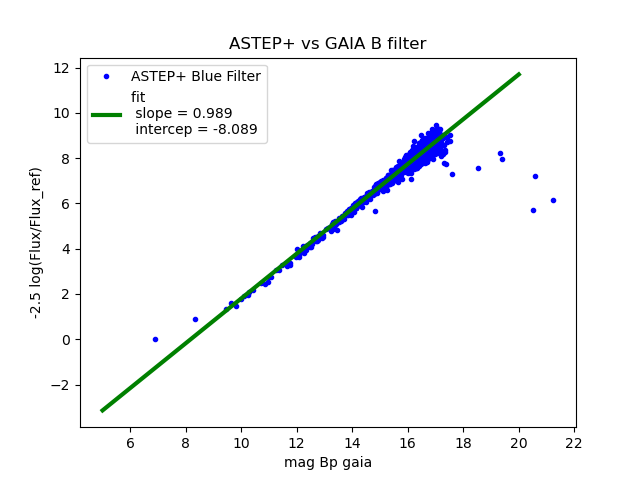}
\includegraphics[height=6.2cm]{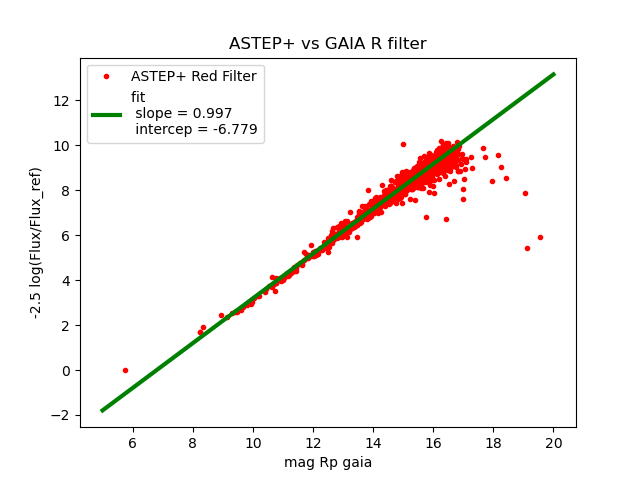}
\end{tabular}
\end{center} 
\caption  {Comparison of the color magnitude B (resp. R) measured on stars of the field of WASP131 as compared to the value B (resp. R) of the same stars in the GAIA catalogue. \label{fig:GAIA_magnitude} } 
\end{figure}

One of the important characterization of the new focal box consists in the comparison between the theoretical transmission of its two photometric channels with other photometric measurements, particularly from space. In Figure \ref{fig:transmissionASTEP}, we present the response of the two-color channels together with the estimated response of the GAIA filters. The transmissions of the blue and red channels of the new focal box are almost identical to the GAIA blue and red channels respectively, although both channels are slightly redder. The transmission of the GAIA red filter is also much higher. 
Figure \ref{fig:GAIA_magnitude} shows the B and R flux of the stars in the ASTEP field (here the WASP 131 field) normalised to the brightest star of the field as compared to the GAIA B and R magnitudes. It shows an almost unity slope of the flux to the magnitude, so without biased estimation, and with a different vertical offset due to the difference in transmission.

\begin{figure}
\begin{center}
\begin{tabular}{c}
\includegraphics[height=7.5cm]{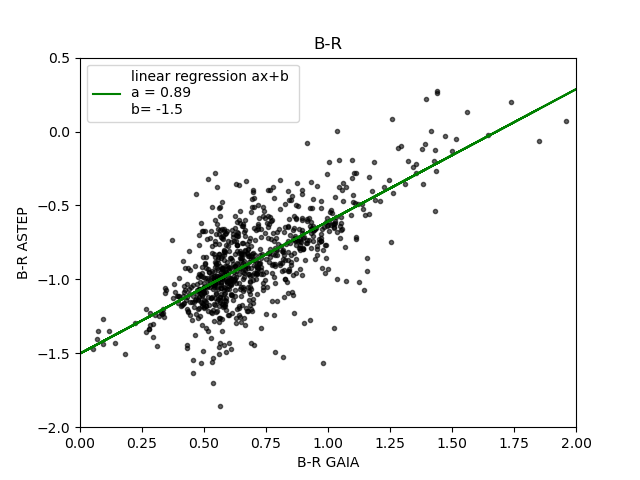}
\end{tabular}
\end{center} 
\caption  {B-R ASTEP+ magnitude computed on the same field and compared to the B-R magnitudes of the stars from the GAIA catalogue. \label{fig:B-R_magnitude} } 
\end{figure}

To be more precise in the color magnitude determination, figure \ref{fig:B-R_magnitude} shows a comparison between B-R measured on all stars of the field of WASP131 using SExtractor software \cite{SExtractor2010} to the color magnitude B-R of the same stars in the GAIA catalogue.The B-R color response of the new focal box to stars of different effective temperatures is also almost the same as that of GAIA, with a different zero point. The magnitude difference between ASTEP+ B-R and GAIA is estimated to 1.5 magnitude.
%The red channel is also very close to the GAIA red channel, although slightly redder is with a reduced transmission. 
% In Section \ref{sec:results}, we show the response of the focal box to stars of different effective temperatures as compared to GAIA, and show that there the response is almost the same than GAIA, with a different zero point. In principle, it would be possible to deduce absolute magnitude and colors from ASTEP+ measurements by comparison with GAIA measurement after calibration of the telescope and atmospheric zero point.

After a period of tests and adjustment, normal operations started at the end of March 2022. An usual observing sequence permits to follow the field around a given star (usually a star known to exhibit transiting exoplanets) for a period of several hours. These observations permit to confirm the parameters of the transit (depth, duration, mean time) and to exclude false detection due to background binaries. It also allows the determination of updated ephemerides for future observations by space mission. When close to the winter solstice, the observations could be almost continuous, allowing the observation of very long transits.

\begin{figure}
\begin{center}
\begin{tabular}{c}
\includegraphics[height=8.5cm]{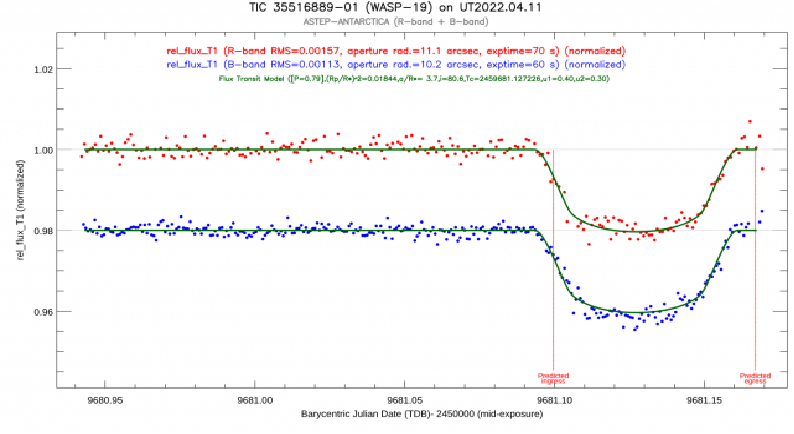}
\end{tabular}
\end{center} 
\caption  {Example of a two color photometric transit of exoplanet WASP-19b recorded at Dome C with ASTEP+ on 2022-04-11. The red points correspond to the data obtained in the ASTEP+ red channel. The blue points correspond to the data obtained in the ASTEP+ blue channel. Note that the transit is achromatic (as expected), and it occurs $\sim 7$\,mns earlier than expected, a consequence of the accumulation of errors on the ephemerides. \label{fig:example} } 
\end{figure}

Figure~\ref{fig:example} shows an example of lightcurves in two colors obtained with the new focal box. Here, we show the transit of WASP-19b that we already observed in 2013 with a single camera\cite{abe2013}. The transit has been recorded in two colors, blue and red, with a similar precision. The dispersion of the measurement is of 0.80\,ppt (parts per thousand) and 1.19\,ppt for an integration of 2 minutes in the Blue and in the Red channels respectively. This has to be compared to the previous dispersion of 1.7 ppt on the same star observed in 2021 with the previous camera. It can be seen that the transit is slightly in advance with respect to the known ephemerides, showing the need for follow-up with instruments such as ASTEP+ to refine the ephemerides for future space missions and to determine possible Transit Timing Variations (TTV) due to the presence of other planets in the system.

\section{CONCLUSION}
A new two-color focal box was successfully installed on the ASTEP telescope at Concordia station for the follow-up of transiting exoplanets. Despite a failure in the communication with one of the camera, the installation was realized without major issues. The new focal box is heavier than the previous one, so we could fear that the guiding could be problematic, as the mount had shown problems in the past. To the contrary, the telescope behaviour and the first results show no problems with the guiding. The quality of the data, even with one camera, is significantly improved with respect to previous observations. With two cameras simultaneously, the determination of transits would be more precise

The cause of the communication issue with the blue camera has been identified and will be solved by next year, allowing a complete use of the focal box. The comparison of the color magnitudes of ASTEP+ with GAIA shows an excellent correlation and allows to determine absolute magnitudes from ASTEP+ data. 

In the future, we will also change the telescope mount for a more powerful one, without gears, allowing faster pointing and guiding which should increase again the performance and allow other type of observing programs. This new mount should be installed in 2024.

%\appendix    %>>>> this command starts appendixes

\section*{ACKNOWLEDGMENTS}       
 
ASTEP and ASTEP+ have benefited from the support of the French and Italian polar agencies IPEV and PNRA, and by the Universit\'e C\^ote d'Azur through Idex UCAJEDI (ANR-15-IDEX-01). We acknowledge support from the European Space Agency (ESA) through the Science Faculty of the European Space Research and Technology Centre (ESTEC). The realisation and installation of the new ASTEP+ focal box was possible thanks to the support of ESA ESTEC and in part funded by the European Union's Horizon 2020 research and innovation programme (grants agreements n$^{\circ}$ 803193/BEBOP), from the Science and Technology Facilities Council (STFC; grant n$^\circ$ ST/S00193X/1), from INSU, and from the laboratoire Lagrange (CNRS UMR 7293). 

% References
\bibliography{report} % bibliography data in report.bib

\begin{thebibliography}{10}

\bibitem{2020SPIE}
{Crouzet}, N., {Agabi}, A., {Guillot}, T., {Abe}, L., {Schmider}, F.-X.,
  {M{\'e}karnia}, D., {Triaud}, A. H.~M.~J., {Bresson}, Y., {Mauclert}, N.,
  {Bailet}, C., {Breeveld}, D., {Blommaert}, S., {Shortt}, B., {Daban}, J.-B.,
  {Lagrange}, A.-M., {Touz{\'e}}, R., {Dufour}, J., {Stee}, V., and {Caruana},
  J., ``{Towards ASTEP+, a two-color photometric telescope at Dome C,
  Antarctica},'' in [{\em Society of Photo-Optical Instrumentation Engineers
  (SPIE) Conference Series}{\nolinebreak\hspace{0.1em}]},  {\em Society of
  Photo-Optical Instrumentation Engineers (SPIE) Conference Series} {\bf
  11447},  114470O (Dec. 2020).

\bibitem{Fressin+2007}
{Fressin}, F., {Guillot}, T., {Schmider}, F.~X., {Agabi}, A., {Moutou}, C.,
  {Aigrain}, S., {Bouchy}, F., {Boer}, M., {Pont}, F., {Erikson}, A., {Rauer},
  H., and {A Step Team}, ``{A STEP: Towards a Large Photometric Survey for
  Exoplanets at Dome C},'' in [{\em EAS Publications
  Series}{\nolinebreak\hspace{0.1em}]},  {Epchtein}, N. and {Candidi}, M.,
  eds., {\em EAS Publications Series} {\bf 25},  225--232 (Jan. 2007).

\bibitem{Daban+2010}
{Daban}, J.-B., {Gouvret}, C., {Guillot}, T., {Agabi}, A., {Crouzet}, N.,
  {Rivet}, J.-P., {Mekarnia}, D., {Abe}, L., {Bondoux}, E.,
  {Fante{\"\i}-Caujolle}, Y., {Fressin}, F., {Schmider}, F.-X., {Valbousquet},
  F., {Blanc}, P.-E., {Le van Suu}, A., {Rauer}, H., {Erikson}, A., {Pont}, F.,
  and {Aigrain}, S., ``{ASTEP 400: a telescope designed for exoplanet transit
  detection from Dome C, Antarctica},'' in [{\em Ground-based and Airborne
  Telescopes III}{\nolinebreak\hspace{0.1em}]},  {Stepp}, L.~M., {Gilmozzi},
  R., and {Hall}, H.~J., eds., {\em Society of Photo-Optical Instrumentation
  Engineers (SPIE) Conference Series} {\bf 7733},  77334T (July 2010).

\bibitem{Guillot+2015}
{Guillot}, T., {Abe}, L., {Agabi}, A., {Rivet}, J.~P., {Daban}, J.~B.,
  {M{\'e}karnia}, D., {Aristidi}, E., {Schmider}, F.~X., {Crouzet}, N.,
  {Gon{\c{c}}alves}, I., {Gouvret}, C., {Ottogalli}, S., {Faradji}, H.,
  {Blanc}, P.~E., {Bondoux}, E., and {Valbousquet}, F., ``{Thermalizing a
  telescope in Antarctica - analysis of ASTEP observations},'' {\em
  Astronomische Nachrichten}~{\bf 336},  638 (Sept. 2015).

\bibitem{Bouma+2020AJ}
{Bouma}, L.~G., {Hartman}, J.~D., {Brahm}, R., {Evans}, P., {Collins}, K.~A.,
  {Zhou}, G., {Sarkis}, P., {Quinn}, S.~N., {de Leon}, J., {Livingston}, J.,
  {Bergmann}, C., {Stassun}, K.~G., {Bhatti}, W., {Winn}, J.~N., {Bakos},
  G.~{\'A}., {Abe}, L., {Crouzet}, N., {Dransfield}, G., {Guillot}, T.,
  {Marie-Sainte}, W., {M{\'e}karnia}, D., {Triaud}, A.~H.~M.~J., {Tinney},
  C.~G., {Henning}, T., {Espinoza}, N., {Jord{\'a}n}, A., {Barbieri}, M.,
  {Nandakumar}, S., {Trifonov}, T., {Vines}, J.~I., {Vuckovic}, M., {Ziegler},
  C., {Law}, N., {Mann}, A.~W., {Ricker}, G.~R., {Vanderspek}, R., {Seager},
  S., {Jenkins}, J.~M., {Burke}, C.~J., {Dragomir}, D., {Levine}, A.~M.,
  {Quintana}, E.~V., {Rodriguez}, J.~E., {Smith}, J.~C., and {Wohler}, B.,
  ``{Cluster Difference Imaging Photometric Survey. II. TOI 837: A Young
  Validated Planet in IC 2602},'' {\em \aj}~{\bf 160},  239 (Nov. 2020).

\bibitem{Dawson+2021}
{Dawson}, R.~I., {Huang}, C.~X., {Brahm}, R., {Collins}, K.~A., {Hobson},
  M.~J., {Jord{\'a}n}, A., {Dong}, J., {Korth}, J., {Trifonov}, T., {Abe}, L.,
  {Agabi}, A., {Bruni}, I., {Butler}, R.~P., {Barbieri}, M., {Collins}, K.~I.,
  {Conti}, D.~M., {Crane}, J.~D., {Crouzet}, N., {Dransfield}, G., {Evans}, P.,
  {Espinoza}, N., {Gan}, T., {Guillot}, T., {Henning}, T., {Lissauer}, J.~J.,
  {Jensen}, E. L.~N., {Sainte}, W.~M., {M{\'e}karnia}, D., {Myers}, G.,
  {Nandakumar}, S., {Relles}, H.~M., {Sarkis}, P., {Torres}, P., {Shectman},
  S., {Schmider}, F.-X., {Shporer}, A., {Stockdale}, C., {Teske}, J., {Triaud},
  A. H.~M.~J., {Wang}, S.~X., {Ziegler}, C., {Ricker}, G., {Vanderspek}, R.,
  {Latham}, D.~W., {Seager}, S., {Winn}, J., {Jenkins}, J.~M., {Bouma}, L.~G.,
  {Burt}, J.~A., {Charbonneau}, D., {Levine}, A.~M., {McDermott}, S., {McLean},
  B., {Rose}, M.~E., {Vanderburg}, A., and {Wohler}, B., ``{Precise Transit and
  Radial-velocity Characterization of a Resonant Pair: The Warm Jupiter
  TOI-216c and Eccentric Warm Neptune TOI-216b},'' {\em \aj}~{\bf 161},  161
  (Apr. 2021).

\bibitem{Dong+2021}
{Dong}, J., {Huang}, C.~X., {Dawson}, R.~I., {Foreman-Mackey}, D., {Collins},
  K.~A., {Quinn}, S.~N., {Lissauer}, J.~J., {Beatty}, T., {Quarles}, B., {Sha},
  L., {Shporer}, A., {Guo}, Z., {Kane}, S.~R., {Abe}, L., {Barkaoui}, K.,
  {Benkhaldoun}, Z., {Brahm}, R., {Bouchy}, F., {Carmichael}, T.~W., {Collins},
  K.~I., {Conti}, D.~M., {Crouzet}, N., {Dransfield}, G., {Evans}, P., {Gan},
  T., {Ghachoui}, M., {Gillon}, M., {Grieves}, N., {Guillot}, T., {Hellier},
  C., {Jehin}, E., {Jensen}, E. L.~N., {Jord{\'a}n}, A., {Kamler}, J.,
  {Kielkopf}, J.~F., {M{\'e}karnia}, D., {Nielsen}, L.~D., {Pozuelos}, F.~J.,
  {Radford}, D.~J., {Schmider}, F.-X., {Schwarz}, R.~P., {Stockdale}, C.,
  {Tan}, T.-G., {Timmermans}, M., {Triaud}, A. H.~M.~J., {Wang}, G., {Ricker},
  G., {Vanderspek}, R., {Latham}, D.~W., {Seager}, S., {Winn}, J.~N.,
  {Jenkins}, J.~M., {Mireles}, I., {Yahalomi}, D.~A., {Morgan}, E.~H., {Vezie},
  M., {Quintana}, E.~V., {Rose}, M.~E., {Smith}, J.~C., and {Shiao}, B.,
  ``{Warm Jupiters in TESS Full-frame Images: A Catalog and Observed
  Eccentricity Distribution for Year 1},'' {\em \apjs}~{\bf 255},  6 (July
  2021).

\bibitem{Grieves+2021}
{Grieves}, N., {Bouchy}, F., {Lendl}, M., {Carmichael}, T., {Mireles}, I.,
  {Shporer}, A., {McLeod}, K.~K., {Collins}, K.~A., {Brahm}, R., {Stassun},
  K.~G., {Gill}, S., {Bouma}, L.~G., {Guillot}, T., {Cointepas}, M., {Dos
  Santos}, L.~A., {Casewell}, S.~L., {Jenkins}, J.~M., {Henning}, T.,
  {Nielsen}, L.~D., {Psaridi}, A., {Udry}, S., {S{\'e}gransan}, D., {Eastman},
  J.~D., {Zhou}, G., {Abe}, L., {Agabi}, A., {Bakos}, G., {Charbonneau}, D.,
  {Collins}, K.~I., {Colon}, K.~D., {Crouzet}, N., {Dransfield}, G., {Evans},
  P., {Goeke}, R.~F., {Hart}, R., {Irwin}, J.~M., {Jensen}, E. L.~N.,
  {Jord{\'a}n}, A., {Kielkopf}, J.~F., {Latham}, D.~W., {Marie-Sainte}, W.,
  {M{\'e}karnia}, D., {Nelson}, P., {Quinn}, S.~N., {Radford}, D.~J.,
  {Rodriguez}, D.~R., {Rowden}, P., {Schmider}, F.-X., {Schwarz}, R.~P.,
  {Smith}, J.~C., {Stockdale}, C., {Suarez}, O., {Tan}, T.-G., {Triaud}, A.
  H.~M.~J., {Waalkes}, W., and {Wingham}, G., ``{Populating the brown dwarf and
  stellar boundary: Five stars with transiting companions near the
  hydrogen-burning mass limit},'' {\em \aap}~{\bf 652},  A127 (Aug. 2021).

\bibitem{Burt+2021}
{Burt}, J.~A., {Dragomir}, D., {Molli{\`e}re}, P., {Youngblood}, A.,
  {Garc{\'\i}a Mu{\~n}oz}, A., {McCann}, J., {Kreidberg}, L., {Huang}, C.~X.,
  {Collins}, K.~A., {Eastman}, J.~D., {Abe}, L., {Almenara}, J.~M.,
  {Crossfield}, I. J.~M., {Ziegler}, C., {Rodriguez}, J.~E., {Mamajek}, E.~E.,
  {Stassun}, K.~G., {Halverson}, S.~P., {Villanueva}, S., {Butler}, R.~P.,
  {Wang}, S.~X., {Schwarz}, R.~P., {Ricker}, G.~R., {Vanderspek}, R., {Latham},
  D.~W., {Seager}, S., {Winn}, J.~N., {Jenkins}, J.~M., {Agabi}, A., {Bonfils},
  X., {Ciardi}, D., {Cointepas}, M., {Crane}, J.~D., {Crouzet}, N.,
  {Dransfield}, G., {Feng}, F., {Furlan}, E., {Guillot}, T., {Gupta}, A.~F.,
  {Howell}, S.~B., {Jensen}, E. L.~N., {Law}, N., {Mann}, A.~W.,
  {Marie-Sainte}, W., {Matson}, R.~A., {Matthews}, E.~C., {M{\'e}karnia}, D.,
  {Pepper}, J., {Scott}, N., {Shectman}, S.~A., {Schlieder}, J.~E., {Schmider},
  F.-X., {Stevens}, D.~J., {Teske}, J.~K., {Triaud}, A. H.~M.~J.,
  {Charbonneau}, D., {Berta-Thompson}, Z.~K., {Burke}, C.~J., {Daylan}, T.,
  {Barclay}, T., {Wohler}, B., and {Brasseur}, C.~E., ``{TOI-1231 b: A
  Temperate, Neptune-sized Planet Transiting the Nearby M3 Dwarf NLTT 24399},''
  {\em \aj}~{\bf 162},  87 (Sept. 2021).

\bibitem{Kaye+2022}
{Kaye}, L., {Vissapragada}, S., {G{\"u}nther}, M.~N., {Aigrain}, S.,
  {Mikal-Evans}, T., {Jensen}, E. L.~N., {Parviainen}, H., {Pozuelos}, F.~J.,
  {Abe}, L., {Acton}, J.~S., {Agabi}, A., {Alves}, D.~R., {Anderson}, D.~R.,
  {Armstrong}, D.~J., {Barkaoui}, K., {Barrag{\'a}n}, O., {Benneke}, B.,
  {Boyd}, P.~T., {Brahm}, R., {Bruni}, I., {Bryant}, E.~M., {Burleigh}, M.~R.,
  {Casewell}, S.~L., {Ciardi}, D., {Cloutier}, R., {Collins}, K.~A., {Collins},
  K.~I., {Conti}, D.~M., {Crossfield}, I. J.~M., {Crouzet}, N., {Daylan}, T.,
  {Dragomir}, D., {Dransfield}, G., {Fabrycky}, D., {Fausnaugh}, M.,
  {Fu{\H{u}}r{\'e}sz}, G., {Gan}, T., {Gill}, S., {Gillon}, M., {Goad}, M.~R.,
  {Gorjian}, V., {Greklek-McKeon}, M., {Guerrero}, N., {Guillot}, T., {Jehin},
  E., {Jenkins}, J.~S., {Lendl}, M., {Kamler}, J., {Kane}, S.~R., {Kielkopf},
  J.~F., {Kunimoto}, M., {Marie-Sainte}, W., {McCormac}, J., {M{\'e}karnia},
  D., {Morales}, F.~Y., {Moyano}, M., {Palle}, E., {Parmentier}, V., {Relles},
  H.~M., {Schmider}, F.-X., {Schwarz}, R.~P., {Seager}, S., {Smith}, A. M.~S.,
  {Tan}, T.-G., {Taylor}, J., {Triaud}, A. H.~M.~J., {Twicken}, J.~D., {Udry},
  S., {Vines}, J.~I., {Wang}, G., {Wheatley}, P.~J., and {Winn}, J.~N.,
  ``{Transit timings variations in the three-planet system: TOI-270},'' {\em
  \mnras}~{\bf 510},  5464--5485 (Mar. 2022).

\bibitem{Wilson+2022}
{Wilson}, T.~G., {Goffo}, E., {Alibert}, Y., {Gandolfi}, D., {Bonfanti}, A.,
  {Persson}, C.~M., {Collier Cameron}, A., {Fridlund}, M., {Fossati}, L.,
  {Korth}, J., {Benz}, W., {Deline}, A., {Flor{\'e}n}, H.-G., {Guterman}, P.,
  {Adibekyan}, V., {Hooton}, M.~J., {Hoyer}, S., {Leleu}, A., {Mustill}, A.~J.,
  {Salmon}, S., {Sousa}, S.~G., {Suarez}, O., {Abe}, L., {Agabi}, A., {Alonso},
  R., {Anglada}, G., {Asquier}, J., {B{\'a}rczy}, T., {Barrado Navascues}, D.,
  {Barros}, S. C.~C., {Baumjohann}, W., {Beck}, M., {Beck}, T., {Billot}, N.,
  {Bonfils}, X., {Brandeker}, A., {Broeg}, C., {Bryant}, E.~M., {Burleigh},
  M.~R., {Buttu}, M., {Cabrera}, J., {Charnoz}, S., {Ciardi}, D.~R.,
  {Cloutier}, R., {Cochran}, W.~D., {Collins}, K.~A., {Col{\'o}n}, K.~D.,
  {Crouzet}, N., {Csizmadia}, S., {Davies}, M.~B., {Deleuil}, M., {Delrez}, L.,
  {Demangeon}, O., {Demory}, B.-O., {Dragomir}, D., {Dransfield}, G.,
  {Ehrenreich}, D., {Erikson}, A., {Fortier}, A., {Gan}, T., {Gill}, S.,
  {Gillon}, M., {Gnilka}, C.~L., {Grieves}, N., {Grziwa}, S., {G{\"u}del}, M.,
  {Guillot}, T., {Haldemann}, J., {Heng}, K., {Horne}, K., {Howell}, S.~B.,
  {Isaak}, K.~G., {Jenkins}, J.~M., {Jensen}, E. L.~N., {Kiss}, L.,
  {Lacedelli}, G., {Lam}, K., {Laskar}, J., {Latham}, D.~W., {Lecavelier des
  Etangs}, A., {Lendl}, M., {Lester}, K.~V., {Levine}, A.~M., {Livingston}, J.,
  {Lovis}, C., {Luque}, R., {Magrin}, D., {Marie-Sainte}, W., {Maxted}, P.
  F.~L., {Mayo}, A.~W., {McLean}, B., {Mecina}, M., {M{\'e}karnia}, D.,
  {Nascimbeni}, V., {Nielsen}, L.~D., {Olofsson}, G., {Osborn}, H.~P.,
  {Osborne}, H. L.~M., {Ottensamer}, R., {Pagano}, I., {Pall{\'e}}, E.,
  {Peter}, G., {Piotto}, G., {Pollacco}, D., {Queloz}, D., {Ragazzoni}, R.,
  {Rando}, N., {Rauer}, H., {Redfield}, S., {Ribas}, I., {Ricker}, G.~R.,
  {Rieder}, M., {Santos}, N.~C., {Scandariato}, G., {Schmider}, F.-X.,
  {Schwarz}, R.~P., {Scott}, N.~J., {Seager}, S., {S{\'e}gransan}, D.,
  {Serrano}, L.~M., {Simon}, A.~E., {Smith}, A. M.~S., {Steller}, M.,
  {Stockdale}, C., {Szab{\'o}}, G., {Thomas}, N., {Ting}, E.~B., {Triaud}, A.
  H.~M.~J., {Udry}, S., {Van Eylen}, V., {Van Grootel}, V., {Vanderspek},
  R.~K., {Viotto}, V., {Walton}, N., and {Winn}, J.~N., ``{A pair of
  sub-Neptunes transiting the bright K-dwarf TOI-1064 characterized with
  CHEOPS},'' {\em \mnras}~{\bf 511},  1043--1071 (Mar. 2022).

\bibitem{Mann+2022}
{Mann}, A.~W., {Wood}, M.~L., {Schmidt}, S.~P., {Barber}, M.~G., {Owen}, J.~E.,
  {Tofflemire}, B.~M., {Newton}, E.~R., {Mamajek}, E.~E., {Bush}, J.~L.,
  {Mace}, G.~N., {Kraus}, A.~L., {Thao}, P.~C., {Vanderburg}, A., {Llama}, J.,
  {Johns-Krull}, C.~M., {Prato}, L., {Stahl}, A.~G., {Tang}, S.-Y., {Fields},
  M.~J., {Collins}, K.~A., {Collins}, K.~I., {Gan}, T., {Jensen}, E. L.~N.,
  {Kamler}, J., {Schwarz}, R.~P., {Furlan}, E., {Gnilka}, C.~L., {Howell},
  S.~B., {Lester}, K.~V., {Owens}, D.~A., {Suarez}, O., {Mekarnia}, D.,
  {Guillot}, T., {Abe}, L., {Triaud}, A. H.~M.~J., {Johnson}, M.~C., {Milburn},
  R.~P., {Rizzuto}, A.~C., {Quinn}, S.~N., {Kerr}, R., {Ricker}, G.~R.,
  {Vanderspek}, R., {Latham}, D.~W., {Seager}, S., {Winn}, J.~N., {Jenkins},
  J.~M., {Guerrero}, N.~M., {Shporer}, A., {Schlieder}, J.~E., {McLean}, B.,
  and {Wohler}, B., ``{TESS Hunt for Young and Maturing Exoplanets (THYME). VI.
  An 11 Myr Giant Planet Transiting a Very-low-mass Star in Lower Centaurus
  Crux},'' {\em \aj}~{\bf 163},  156 (Apr. 2022).

\bibitem{Christian+2022}
{Christian}, S., {Vanderburg}, A., {Becker}, J., {Yahalomi}, D.~A., {Pearce},
  L., {Zhou}, G., {Collins}, K.~A., {Kraus}, A.~L., {Stassun}, K.~G., {de
  Beurs}, Z., {Ricker}, G.~R., {Vanderspek}, R.~K., {Latham}, D.~W., {Winn},
  J.~N., {Seager}, S., {Jenkins}, J.~M., {Abe}, L., {Agabi}, K., {Amado},
  P.~J., {Baker}, D., {Barkaoui}, K., {Benkhaldoun}, Z., {Benni}, P.,
  {Berberian}, J., {Berlind}, P., {Bieryla}, A., {Esparza-Borges}, E., {Bowen},
  M., {Brown}, P., {Buchhave}, L.~A., {Burke}, C.~J., {Buttu}, M., {Cadieux},
  C., {Caldwell}, D.~A., {Charbonneau}, D., {Chazov}, N., {Chimaladinne}, S.,
  {Collins}, K.~I., {Combs}, D., {Conti}, D.~M., {Crouzet}, N., {de Leon},
  J.~P., {Deljookorani}, S., {Diamond}, B., {Doyon}, R., {Dragomir}, D.,
  {Dransfield}, G., {Essack}, Z., {Evans}, P., {Fukui}, A., {Gan}, T.,
  {Esquerdo}, G.~A., {Gillon}, M., {Girardin}, E., {Guerra}, P., {Guillot}, T.,
  {K. Habich}, E.~K., {Henriksen}, A., {Hoch}, N., {Isogai}, K.~I., {Jehin},
  E., {Jensen}, E. L.~N., {Johnson}, M.~C., {Livingston}, J.~H., {Kielkopf},
  J.~F., {Kim}, K., {Kawauchi}, K., {Krushinsky}, V., {Kunzle}, V., {Laloum},
  D., {Leger}, D., {Lewin}, P., {Mallia}, F., {Massey}, B., {Mori}, M.,
  {McLeod}, K.~K., {M{\'e}karnia}, D., {Mireles}, I., {Mishevskiy}, N.,
  {Tamura}, M., {Murgas}, F., {Narita}, N., {Naves}, R., {Nelson}, P.,
  {Osborn}, H.~P., {Palle}, E., {Parviainen}, H., {Plavchan}, P., {Pozuelos},
  F.~J., {Rabus}, M., {Relles}, H.~M., {Rodr{\'\i}guez L{\'o}pez}, C., {Quinn},
  S.~N., {Schmider}, F.-X., {Schlieder}, J.~E., {Schwarz}, R.~P., {Shporer},
  A., {Sibbald}, L., {Srdoc}, G., {Stibbards}, C., {Stickler}, H., {Suarez},
  O., {Stockdale}, C., {Tan}, T.-G., {Terada}, Y., {Triaud}, A., {Tronsgaard},
  R., {Waalkes}, W.~C., {Wang}, G., {Watanabe}, N., {Wenceslas}, M.-S.,
  {Wingham}, G., {Wittrock}, J., and {Ziegler}, C., ``{A Possible Alignment
  Between the Orbits of Planetary Systems and their Visual Binary
  Companions},'' {\em \aj}~{\bf 163},  207 (May 2022).

\bibitem{Dransfield+2022}
{Dransfield}, G., {Triaud}, A. H.~M.~J., {Guillot}, T., {Mekarnia}, D.,
  {Nesvorn{\'y}}, D., {Crouzet}, N., {Abe}, L., {Agabi}, K., {Buttu}, M.,
  {Cabrera}, J., {Gandolfi}, D., {G{\"u}nther}, M.~N., {Rodler}, F.,
  {Schmider}, F.-X., {Stee}, P., {Suarez}, O., {Collins}, K.~A.,
  {D{\'e}vora-Pajares}, M., {Howell}, S.~B., {Matthews}, E.~C., {Standing},
  M.~R., {Stassun}, K.~G., {Stockdale}, C., {Quinn}, S.~N., {Ziegler}, C.,
  {Crossfield}, I. J.~M., {Lissauer}, J.~J., {Mann}, A.~W., {Matson}, R.,
  {Schlieder}, J., and {Zhou}, G., ``{HD 28109 hosts a trio of transiting
  Neptunian planets including a near-resonant pair, confirmed by ASTEP from
  Antarctica},'' {\em \mnras}  (May 2022).

\bibitem{2022SPIE+}
{Dransfield}, G., {Mekarnia}, D., {Triaud}, A.~H., {Guillot}, T., {Abe}, L.,
  {Timmermans}, M., {Garcia}, L., {Crouzet}, N., {Schmider}, F.-X., {Agabi},
  A., {Suarez}, O., {Bendjoya}, P., {Günther}, M., and {Lai}, O.,
  ``{Observation Scheduling and Automatic Data Reduction for the Antarctic
  ASTEP+ telescope},'' in [{\em Society of Photo-Optical Instrumentation
  Engineers (SPIE) Conference Series}{\nolinebreak\hspace{0.1em}]},   this
  proceedings (2022).

\bibitem{Lai2019}
{Lai}, O., {Withington}, J.~K., {Laugier}, R., and {Chun}, M., ``{Direct
  measure of dome seeing with a localized optical turbulence sensor},'' {\em
  MNRAS}~{\bf 484},  5568--5577 (Apr. 2019).

\bibitem{SExtractor2010}
Bertin, E. and Arnouts, S., ``Sextractor: Source extractor,'' {\em Astrophysics
  Source Code Library} ,  10064-- (10 2010).

\bibitem{abe2013}
Abe, L., Gon{\c{c}}alves, I., Agabi, A., Alapini, A., Guillot, T.,
  M{\'e}karnia, D., Rivet, J.-P., Schmider, F.-X., Crouzet, N., Fortney, J.,
  et~al., ``The secondary eclipses of wasp-19b as seen by the astep 400
  telescope from antarctica,'' {\em Astronomy \& Astrophysics}~{\bf 553},  A49
  (2013).

\end{thebibliography}
\bibliographystyle{spiebib} % makes bibtex use spiebib.bst

\end{document}